\documentclass[aps,prb,a4paper,superscriptaddress,twocolumn,floatfix,showpacs]{revtex4}
\usepackage{graphicx}
\usepackage{latexsym}
\usepackage[]{amsmath}
\usepackage{natbib}

\begin{document}
\newcommand{\chem}[1]{\ensuremath{\mathrm{#1}}}

\title{Muon-spin relaxation and heat capacity measurements on the \\
magnetoelectric and multiferroic pyroxenes LiFeSi$_2$O$_6$ and NaFeSi$_2$O$_6$}

\author{P.~J.~Baker}
\email{peter.baker@stfc.ac.uk}
\affiliation{Oxford University Department of Physics, Clarendon Laboratory,
Parks Road, Oxford OX1 3PU, United Kingdom}
\affiliation{ISIS Pulsed Neutron and Muon Source, STFC Rutherford Appleton Laboratory, 
Harwell Science and Innovation Campus, Didcot, Oxfordshire, OX11 0QX, United Kingdom}

\author{H.~J.~Lewtas}
\affiliation{Oxford University Department of Physics, Clarendon Laboratory,
Parks Road, Oxford OX1 3PU, United Kingdom}

\author{S.~J.~Blundell}
\affiliation{Oxford University Department of Physics, Clarendon Laboratory,
Parks Road, Oxford OX1 3PU, United Kingdom}

\author{T.~Lancaster}
\affiliation{Oxford University Department of Physics, Clarendon Laboratory,
Parks Road, Oxford OX1 3PU, United Kingdom}

\author{I.~Franke}
\affiliation{Oxford University Department of Physics, Clarendon Laboratory,
Parks Road, Oxford OX1 3PU, United Kingdom}

\author{W.~Hayes}
\affiliation{Oxford University Department of Physics, Clarendon Laboratory,
Parks Road, Oxford OX1 3PU, United Kingdom}

\author{F.~L.~Pratt}
\affiliation{ISIS Pulsed Neutron and Muon Source, STFC Rutherford Appleton Laboratory, 
Harwell Science and Innovation Campus, Didcot, Oxfordshire, OX11 0QX, United Kingdom}

\author{L.~Bohat\'{y}}
\affiliation{Institut f\"{u}r Kristallographie, Universit\"{a}t zu K\"{o}ln, Z\"{u}lpicher Stra{\ss}e 49 b, 50674, K\"{o}ln, 
Germany}

\author{P.~Becker}
\affiliation{Institut f\"{u}r Kristallographie, Universit\"{a}t zu K\"{o}ln, Z\"{u}lpicher Stra{\ss}e 49 b, 50674, K\"{o}ln, 
Germany}

\date{\today}

\begin{abstract}
The results of muon-spin relaxation and heat capacity measurements on two pyroxene compounds LiFeSi$_2$O$_6$ and NaFeSi$_2$O$_6$ demonstrate that despite their underlying structural similarity the magnetic ordering is 
considerably different. In LiFeSi$_2$O$_6$ a single muon precession frequency is observed below $T_{\rm N}$, consistent with a single peak at $T_{\rm N}$ in the heat capacity and a commensurate magnetic structure. In applied magnetic fields the heat capacity peak splits in two. In contrast, for natural NaFeSi$_2$O$_6$, where multiferroicity has been observed in zero-magnetic-field, a rapid Gaussian depolarization is observed showing that the magnetic structure is more complex. Synthetic NaFeSi$_2$O$_6$ shows a single muon precession frequency but with a far larger damping rate than in the lithium compound.
Heat capacity measurements reproduce the phase diagrams previously derived from other techniques and demonstrate that the magnetic entropy is mostly associated with the build up of correlations in the quasi-one-dimensional Fe$^{3+}$ chains. 
\end{abstract}

\pacs{76.75.+i, 75.50.Ee, 75.85.+t}

\maketitle

\section{\label{sec:intro} Introduction}
Multiferroic materials demonstrating coupled magnetic and ferroelectric order have once again become an active field of research, since they offer both interesting physical properties and the possibility of technological applications.~\cite{fiebig05,tokura06,eerenstein06,spaldin06,khomskii06,cheong07}
While an increasing number of multiferroic materials have been discovered in recent years~\cite{kimura03,hur04,lawes05,heyer06,park07,rusydi08,choi08}
and much progress has been made in finding general rules to describe the origins of this effect~\cite{mostovoy06,betouras07} it is not always possible to predict if a given material will be multiferroic. Isostructural series have already provided considerable insights into multiferroicity, notable examples being the hexagonal and orthorhombic manganites.~\cite{kimura05,lee08} Competing magnetic interactions and a strong magnetoelastic coupling are both known to favour multiferroicity. In this context the discoveries of multiferroicity in the pyroxene compound \chem{NaFeSi_{2}O_{6}} and magnetoelectricity in \chem{LiFeSi_{2}O_{6}} have suggested that this geologically common family may offer more multiferroic compounds, as well as providing an opportunity to study isostructural materials with different spins and magnetic exchange constants.~\cite{jodlauk07}

\begin{figure}[t]
\centering
\includegraphics[width=\columnwidth]{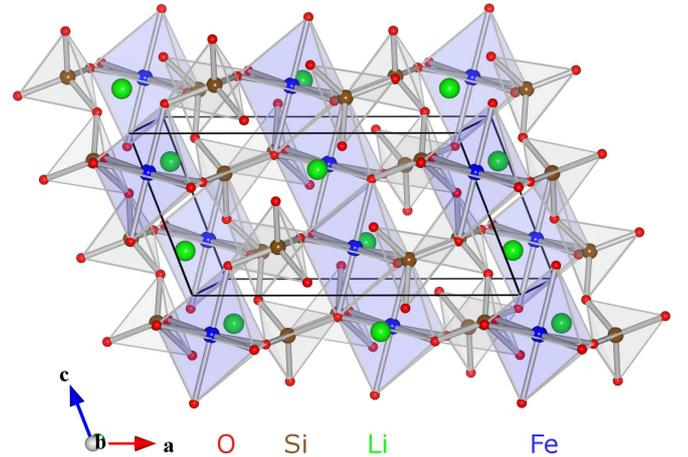}
\caption{(Color online)
Structure of \chem{LiFeSi_{2}O_{6}} showing the \chem{Fe^{3+}} chains running along the $c$-axis linked by \chem{SiO_4} tetrahedra. The structural data come from Ref.~\onlinecite{redhammer09}.
}
\label{fig:structure}
\end{figure}

Pyroxene compounds have chemical formulae \chem{A^{+}M^{3+}(Si,Ge)_{2}O_6} and chains of \chem{M^{3+}} ions surrounded by oxygen octahedra lie along the crystallographic $c$-axis.~\cite{redhammer04} The \chem{M^{3+}} chains 
are connected by \chem{(Si,Ge)O_4} tetrahedra. This structure is shown in Fig.~\ref{fig:structure}.
Most magnetic members of this family show N\'{e}el ordering at low temperature, a notable exception being the orbitally assisted spin-Peierls transition seen in \chem{NaTiSi_{2}O_{6}}.~\cite{isobe02,baker07} 
Common to both the N\'{e}el ordered and spin-gapped compounds is the dominant intrachain exchange interaction giving quasi-one-dimensional magnetic properties.

\chem{LiFeSi_{2}O_6} has a N\'{e}el temperature of $18$~K and there is no pyroelectric current without an applied magnetic field.~\cite{jodlauk07} Applying a magnetic field along the $c$-axis reduces the temperature of the peak 
of the magnetic susceptibility to $14$~K at $14$~T and measurements of the pyroelectric current, $I_b$, show a peak which follows the same magnetic field dependence as that in the magnetic susceptibility. Smaller peaks in $I_b$ at higher temperature were also observed but their origin is unclear.~\cite{jodlauk07} The magnetic structure has been determined to be antiferromagnetically coupled ferromagnetic chains with magnetic space group $P2_{1}/c^{\prime}$.~\cite{redhammer09} 
This magnetic structure allows for magnetoelectric effects consistent with those observed. The isostructural compound \chem{LiCrSi_{2}O_{6}} was found to have comparable magnetic and magnetoelectric properties.~\cite{jodlauk07,nenert09}

The situation in \chem{NaFeSi_{2}O_{6}} is rather more complex, largely because of the differences observed between natural and synthetic samples. Natural samples, which are known to contain impurities, show two phase transitions in zero magnetic field: at $8$~K to a collinear magnetic structure and at $6$~K to a ferroelectric ($P \parallel b$) phase. In fields above $4$~T a ferroelectric ($P \parallel c$) phase was observed below $5$~K.~\cite{jodlauk07} Synthetic samples showed a similar magnetic structure to \chem{LiFeSi_{2}O_{6}}, with antiferromagnetically coupled ferromagnetic chains, albeit with evidence for a further incommensurate modulation to this structure that could not be determined.~\cite{ballet89} Given that it was not possible to index all the magnetic Bragg peaks, the results of {\it ab initio} calculations, and the multiferroicity observed in the natural samples it seems likely that the magnetic ordering is actually helical and incommensurate.~\cite{jodlauk07} 

More detailed {\it ab initio} calculations for a broad range of pyroxene compounds,
including those we study here, were carried out by \textcite{streltsov08}. They modelled the exchange constants in terms of an intrachain exchange $J$, and two interchain exchange constants, $J_1$ and $J_2$, all of which were found to be antiferromagnetic for both compounds. The calculations suggest $J^{\rm Li} = 7$~K, $J^{\rm Li}_1 =1.9$~K, and $J^{\rm Li}_2 =3.4$~K; and $J^{\rm Na} = 8.5$~K, $J^{\rm Na}_1 =0.8$~K, and $J^{\rm Na}_2 = 1.6$~K.~\cite{streltsov08} 
These values suggest that the magnetism in \chem{LiFeSi_{2}O_6} is likely to be more three-dimensional than that in \chem{NaFeSi_{2}O_6} and the different exchange constants may have an even more significant effect on the fine details of the magnetic structure and any magnetostriction.

In this paper we investigate synthetic samples of the two pyroxene compounds \chem{LiFeSi_{2}O_{6}} and \chem{NaFeSi_{2}O_{6}}, and a natural sample of \chem{NaFeSi_{2}O_{6}}, using heat capacity and muon-spin relaxation measurements. These probe the change in magnetic entropy around the phase transitions and the local magnetic field distributions within the samples. While there are some underlying similarities in the magnetic properties, the effects of the changing exchange constants and the presence of impurity-induced disorder in the natural sample are clearly evident in the data recorded by both techniques.

\section{\label{sec:expdetails} Experimental}
\subsection{\label{sec:samples} Samples}
Our natural sample of \chem{NaFeSi_{2}O_{6}} was cut from the same crystal that was used in Ref.~\onlinecite{jodlauk07}. Electron microprobe analysis has shown that the composition is
\chem{Na_{1.04}Fe_{0.83}Ca_{0.04}Mn_{0.02}Al_{0.01}Ti_{0.08}Si_{2}O_6}.~\cite{jodlauk07} The synthetic sample of \chem{LiFeSi_{2}O_{6}} was composed of small translucent single crystals grown from melt solution, see also Ref.~\onlinecite{jodlauk07}. The powder sample of synthetic \chem{NaFeSi_{2}O_{6}} was obtained by crystallisation of glassy \chem{NaFeSi_{2}O_{6}} that was prepared using high-temperature flux.

\subsection{\label{sec:HC} Heat capacity measurements}
\begin{figure}[t]
\includegraphics[width=\columnwidth]{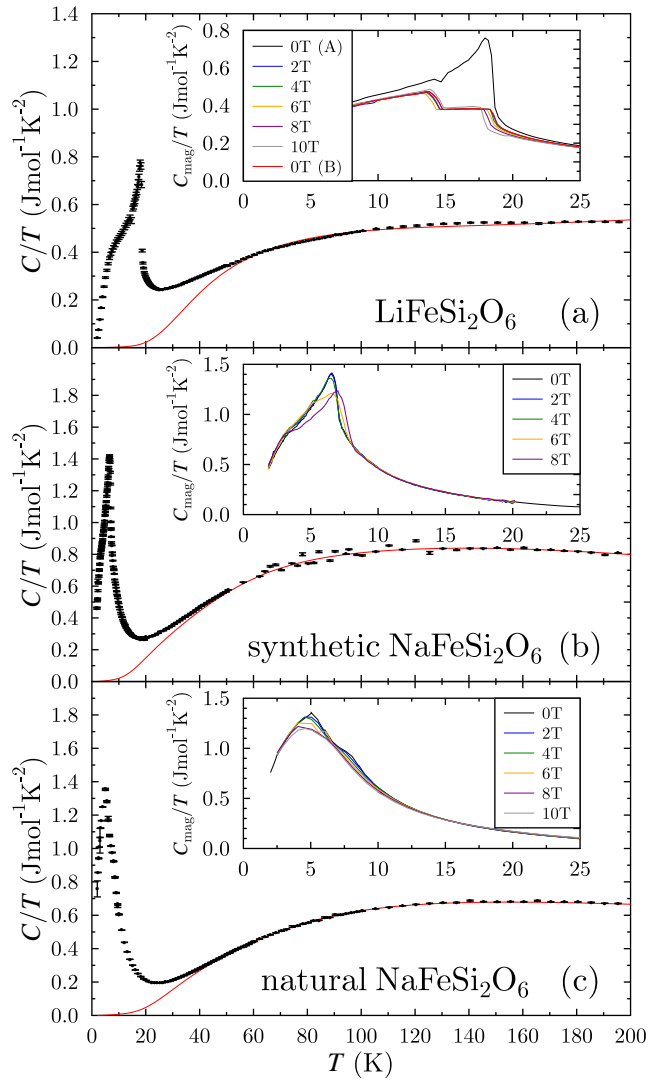}
\caption{ 
(Color online) Heat capacity measurements on 
(a) \chem{LiFeSi_{2}O_{6}},
(b) synthetic \chem{NaFeSi_{2}O_{6}},  
and (c) natural \chem{NaFeSi_{2}O_6}. 
The lines in the main panels show the fitted lattice terms described 
in the text and the insets show the variation with applied magnetic 
field close to the magnetic ordering transitions described in the text. 
\label{hc}}
\end{figure}

Heat capacity measurements were made using a Quantum Design Physical Properties Measurement System (PPMS), employing the two-tau relaxation method, in magnetic fields between 0 and 10~T. The data for both compounds are shown in Fig.~\ref{hc}. They show clear peaks associated with the magnetic ordering transitions found using other techniques.~\cite{jodlauk07} In compounds such as these, a (partial) hump in the heat capacity is observed above $T_{\rm N}$ as the correlations build up, and then a peak is superimposed upon this as the system enters a three-dimensionally ordered state. Some information concerning the dimensionality of the system can be obtained from the form of the hump and the relative size of the peak.~\cite{dejongh74} The lattice contribution to the heat capacity of each compound was modelled using one Debye and two Einstein components. Parameters derived from fitting this form to the data above $50$~K are given in Table~\ref{hctable}. 

In zero applied field \chem{LiFeSi_{2}O_{6}} shows the build up of short-ranged correlations in the chains from well above $T_{\rm N}$ and a single peak in the heat capacity at $T_{\rm N} = 18$~K. 
Having subtracted the lattice contribution, we estimate that the integrated magnetic entropy up to $50$~K is $11.2$~Jmol$^{-1}$K$^{-1}$, with around $80$~\% accounted for by the short-range correlations. 
The heat capacity measurements on \chem{LiFeSi_{2}O_{6}} show an unusual hysteresis with a magnetic field applied perpendicular to the $ab$ plane. Measuring in successively increasing magnetic fields up to 10~T repeated the peak in the data seen in zero-field [0~T (A)] within the experimental error. However, measurements in successively decreasing fields, while cooling the sample from around $2\times T_{\rm N}$, gave the two field dependent steps 
in the heat capacity shown in the inset to Fig.~\ref{hc}(a). No pattern is evident in different positions of these steps in different fields. The two features were accompanied by small amounts of latent heat (evident in the poorer fits to the raw thermal relaxation data recorded by the PPMS) and persist down to zero applied field [0~T (B)]. This behavior suggests short range order persists well above $T_{\rm N}$ and produces hysteresis in the sample when fields are applied.

The heat capacity data on synthetic \chem{NaFeSi_{2}O_{6}} are shown in Fig.~\ref{hc}(b) and take a similar form to the data for \chem{LiFeSi_{2}O_{6}}, with a significant magnetic heat capacity well above $T_{\rm N}$. A clear peak is found at $6.6$~K, rather broader than in \chem{LiFeSi_{2}O_{6}} and a little lower than the transition temperature found in the $\mu$SR measurements. The peak broadens with increasing field but does not move significantly. Our data are in excellent quantitative agreement with those reported previously by \textcite{ko77} for their measurements on a synthetic sample.
The integrated magnetic entropy up to $50$~K is $11.2$~Jmol$^{-1}$K$^{-1}$ with around two-thirds of this appearing to be associated with the build up of correlations within the chains.

In natural \chem{NaFeSi_{2}O_{6}} [Fig.~\ref{hc}(c)] the field-dependent part of the heat capacity forms a much smaller fraction of the feature around $7$~K, which is dominated by the build up of correlations in the chains. Examining the data closely shows two small peaks at $8$~K and $6$~K corresponding to the magnetic transitions 
and these merge as the magnetic field is increased, consistent with the phase diagram proposed in Ref.~\onlinecite{jodlauk07}. The magnetic entropy integrated up to $50$~K is approximately $13.9$~Jmol$^{-1}$K$^{-1}$, approximately 90~\% of which is accounted for by short-ranged ordering. It is notable that the field dependent heat capacity of \chem{LiFeSi_{2}O_{6}} shows two peak features in applied field that are similar to those seen in natural \chem{NaFeSi_{2}O_{6}}. The origin of the two features is at present unknown and neutron scattering experiments in applied field should be carried out on \chem{LiFeSi_{2}O_{6}} to discover whether a magnetic transition is involved.

\begin{table}[t]
\centering
\begin{tabular}{|l|c|c|c|}
\hline
Sample & LiFeSi$_2$O$_6$ & NaFeSi$_2$O$_6$ & NaFeSi$_2$O$_6$ \\
 & synthetic & synthetic & natural \\
\hline
$\theta_{\rm D}$~(K) & 621(18) & 370(20) & 530(20) \\
$\theta_{\rm E1}$~(K) & 192(3) & 190(10) & 165(5) \\
$\theta_{\rm E2}$~(K) & 1210(50) & 700(50) & 1050(50) \\
\hline
\end{tabular}
\caption{Parameters and their statistical errors derived from fitting one Debye and two Einstein components to the heat capacity data above $50$~K.}
\label{hctable}
\end{table}

\subsection{\label{sec:musr} $\mu$SR measurements}
Our positive muon-spin relaxation ($\mu$SR) measurements~\cite{blundell99} 
($\tau_{\mu}=2.2~\mu$s, $\gamma_{\mu} = 2\pi \times 135.5$~MHzT$^{-1}$) were 
carried out on the General Purpose Surface-Muon Instrument (GPS) at the Paul Scherrer Institute, Switzerland. Samples were mounted on a low background sample holder with aluminized mylar tape to minimize the background from muons stopping outside the sample. To measure the time evolution of the muon spin polarization, emitted decay positrons were collected in detectors placed forward (F) and backward (B) relative to the initial muon spin direction (antiparallel to the beam momentum).  The muon decay asymmetry is defined in terms of the count rates in the two detectors ($N_{\rm F}$
and $N_{\rm B}$) as:
\begin{equation}
A(t) = \frac{N_{\rm F}(t) - \alpha N_{\rm B}(t)}{N_{\rm F}(t) 
+ \alpha N_{\rm B}(t)},
\label{asymmetry}
\end{equation}
where $\alpha$ is an experimental calibration constant related to the relative efficiency of the detectors.

\begin{figure}[t]
\includegraphics[width=\columnwidth]{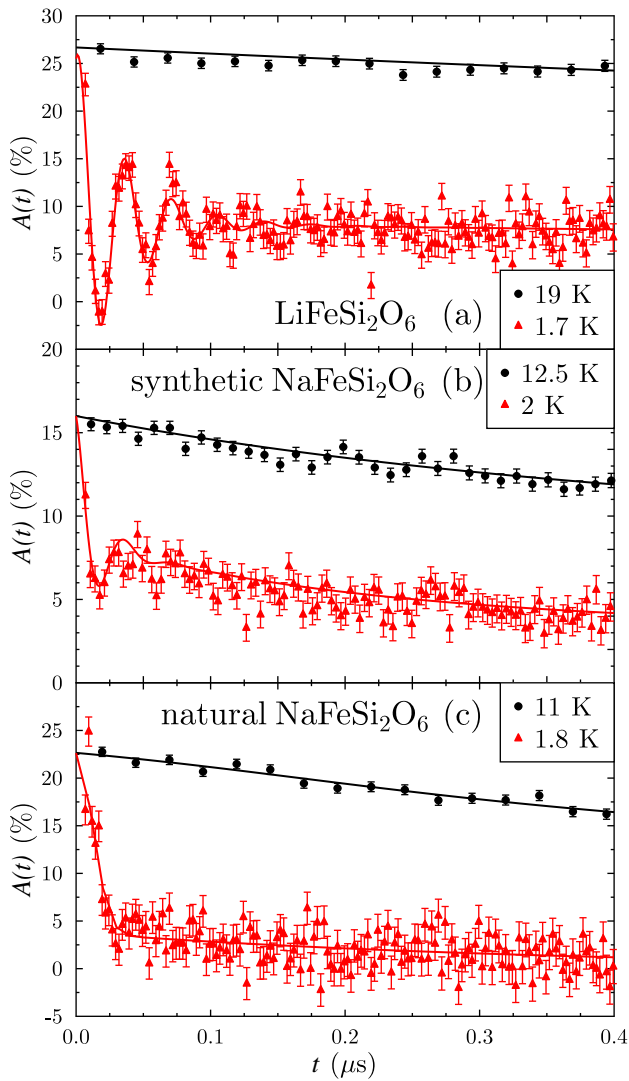}
\caption{
(Color online) 
Muon asymmetry data above and below the magnetic transitions for: 
(a) \chem{LiFeSi_{2}O_{6}}, 
(b) synthetic \chem{NaFeSi_{2}O_{6}}, 
and
(c) natural \chem{NaFeSi_{2}O_{6}}. 
Below $T_{\rm N}$ we fit with Eq.~\ref{lifitfunc} for (a) and (b), and Eq.~\ref{nafitfunc} for (c). Above $T_{\rm N}$ the relaxation is exponential.
The width of the time bins in the asymmetry histograms have been increased for clarity. For the synthetic \chem{NaFeSi_{2}O_6} measurements were done with the initial muon spin rotated differently relative to the detectors, leading to the lower asymmetry values.
\label{datacompare}}
\end{figure}

\begin{figure}[t]
\includegraphics[width=\columnwidth]{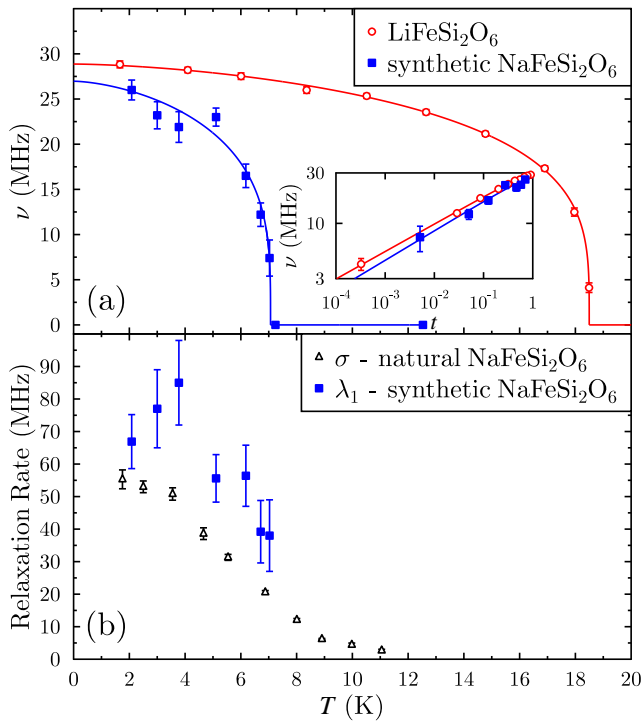}
\caption{
(Color online) 
(a) Muon oscillation frequencies, $\nu$ (Eq.~\ref{lifitfunc}), for \chem{LiFeSi_{2}O_{6}} and the synthetic sample of \chem{NaFeSi_{2}O_6} with fits to Eq.~\ref{nuT} described in the text. 
(Inset) The oscillation frequencies plotted against reduced temperature, $t = (T_{\rm N}-T)/T_{\rm N}$, showing the similarity of the trends approaching $T_{\rm N}$.
(b) The Gaussian relaxation rate, $\sigma$ (Eq.~\ref{nafitfunc}), for the natural \chem{NaFeSi_{2}O_6} and the linewidth $\lambda_1$ for the synthetic \chem{NaFeSi_{2}O_6}. 
\label{muoncompare}}
\end{figure}

The muon spins are sensitive to both static and fluctuating local magnetic
fields at their stopping positions inside the material, and these affect 
how the form of the muon decay asymmetry changes with time. In the 
paramagnetic phase of each compound the muon relaxation is well described 
by a single exponential relaxation.
In \chem{LiFeSi_{2}O_{6}} and synthetic \chem{NaFeSi_{2}O_{6}} [Fig.~\ref{datacompare}~(a) and (b)] we observe coherent muon precession below $T_{\rm N}$ consistent with long range magnetic order and quasistatic magnetic fields at the muon stopping site. The data are well described by the function: 
\begin{equation}
A(t) = A_1 e^{-\lambda_{1} t} \cos(2\pi\nu t)+ A_2 e^{-\lambda_2 t}.
\label{lifitfunc}
\end{equation}
The first term describes damped muon precession around quasistatic local fields ($B = 2\pi\nu / \gamma_{\mu}$) perpendicular to the muon spin polarization and the second term is an exponential relaxation, of rate $\lambda_2$, due to fluctuations flipping the spins of muons having a non-zero spin component along the local magnetic field direction. The values of $\nu$ derived from fitting the asymmetry data are shown in Fig.~\ref{muoncompare}(a). 
We find that about $2/3$ of the asymmetry is associated with the oscillating signal, consistent with the fact that in a polycrystalline sample $2/3$ of the muons will find local magnetic fields perpendicular to their spin polarization and $1/3$ will experience fields along their spin direction that can only lead to depolarization if fluctuations are present.
In \chem{LiFeSi_{2}O_{6}} $\lambda_1$ and $\lambda_2$ are almost temperature-independent. The parameters extracted from the $\mu$SR data analysis are presented in Fig.~\ref{muoncompare}. As shown in Fig.~\ref{muoncompare}(a), the precession frequencies in \chem{LiFeSi_{2}O_{6}} and synthetic \chem{NaFeSi_{2}O_{6}} are well described by the phenomenological function:
\begin{equation} 
\nu(T) = \nu(0)(1-(T/T_{\rm N})^{\alpha})^{\beta}.
\label{nuT}
\end{equation} 
For \chem{LiFeSi_{2}O_6}, $\nu(0) = 28.9(3)$~MHz, $T_{\rm N} = 18.50(1)$~K, $\alpha = 1.6(2)$, and $\beta = 0.26(2)$. This implies that the $T \rightarrow 0$ internal field at the muon site is approximately $0.2$~T. 
For synthetic \chem{NaFeSi_{2}O_{6}} the frequency is less well defined because the oscillations are far more strongly damped and, constraining $\alpha$ to the value found for the Li sample, we can fit $\nu(0) = 27(1)$~MHz, $T_{\rm N} = 7.07(5)$~K, and $\beta = 0.28(4)$.

Data for natural \chem{NaFeSi_{2}O_{6}} takes a different form to that for the other two samples [see Fig.~\ref{datacompare}~(c)]. Here we see no coherent muon precession, suggesting a large range of quasistatic fields occurs at muon stopping sites. A random distribution of static magnetic fields leads to a Kubo-Toyabe function~\cite{blundell99}, which shows a dip and recovery in the asymmetry. Even with the addition of a damping term it is not possible to get reliable fits to the measured data using such a fitting function. Instead, we can effectively describe the measured asymmetry using a rapid Gaussian relaxation to describe the effect of the quasistatic fields and a slow exponential that describes the $1/3$ tail expected for the Kubo-Toyabe function:
\begin{equation}
A(t) = A_1 e^{-\sigma^2 t^2} + A_2 e^{-\lambda t}.
\label{nafitfunc}
\end{equation}
In analogy with Eq.~\ref{lifitfunc} the first term describes the incoherent precession about large static magnetic fields and the second term describes spin-flipping of muons with their spin direction aligned along the local magnetic field. The experiment on natural \chem{NaFeSi_{2}O_6} was carried out on a large single crystal but we have no expectation for the ratio $A_1 : A_2$ because the details of the magnetic structure are unclear. 
In Fig.~\ref{muoncompare}~(b) we present the values of $\sigma$ derived from the asymmetry data for natural \chem{NaFeSi_{2}O_{6}} and the linewidth $\lambda_1$ associated with the oscillating component of the signal in the synthetic sample. The relaxation rate $\sigma$ does not follow the same power law as the precession frequencies, suggesting that the static magnetism does not emerge in as well defined a manner as in the synthetic samples, instead growing smoothly through the two transitions observed previously~\cite{jodlauk07} with no evidence for an intermediate collinearly ordered phase. We can estimate the range of magnetic fields at the muon sites using the relation $\Delta B = \sqrt{2}\sigma/\gamma_{\mu} \sim 0.6$~T, which is larger than the value of $\sim 0.2$~T associated with the precession frequencies in the other two samples. 
To gain a further understanding of this we firstly calculated the dipole field distribution for two plausible model magnetic structures, ferromagnetic chains coupled antiferromagnetically, and antiferromagnetic chains coupled antiferromagnetically. In both cases the moments were taken to lie along the chain direction.
The results from these two calculations are very similar, with the muons sitting near the oxygen atoms linking the Fe octahedra and Si tetrahedra, approximately $a/4$ from the Fe chains.
Because of this we cannot distinguish the magnetic structure.
The second stage was to model the effect of local site dilution at the $\sim 17$~\% of \chem{Fe^{3+}} sites which are not occupied by \chem{Fe^{3+}} ions. In this naive model we assume that all the dopants are non-magnetic and calculated the dipole field of individual ions at muon sites.
The distribution of fields is dominated by the effect of the closest iron moment to the muon site, which is around $\sim 0.35$~T, but averaging over the neighbouring sites leads to a distribution width $\Delta B \sim 0.25$~T.
Canting the moment direction towards the $a$-axis enhances this effect by up to a factor of two, but such a large canting is not consistent with the previous neutron diffraction results.~\cite{ballet89}
On this basis, site dilution alone cannot explain the magnetic field distribution observed in natural \chem{NaFeSi_{2}O_{6}}, but is consistent with the marginal change in the depolarization seen between the synthetic and natural samples. 

\section{\label{sec:discussion} Discussion}
Our heat capacity and $\mu$SR results show that \chem{LiFeSi_{2}O_{6}} is a commensurate antiferromagnet in zero-field, in agreement with the neutron diffraction results reported previously.~\cite{redhammer01} The value of  $\beta=0.26(2)$ extracted from the temperature dependence of the oscillation frequencies suggests the magnetic ordering below $T_{\rm N}$ is intermediate between two- and three-dimensional behavior.
Synthetic \chem{NaFeSi_{2}O_{6}} shows similar heat capacity features to \chem{LiFeSi_{2}O_{6}} but the muon oscillations are heavily damped. This suggests a more complex magnetic ordering where muons stopping at structurally equivalent sites experience a broad range of magnetic fields. The temperature dependence of the precession frequency in the synthetic \chem{NaFeSi_{2}O_{6}} sample is broadly consistent with that seen in the \chem{LiFeSi_{2}O_{6}} sample. This is seen more clearly when plotting the oscillation frequencies in the two samples against reduced temperature, $t = (T_{\rm N}-T)/T_{\rm N}$, shown in the inset to Fig.~\ref{muoncompare}~(a). 
Our heat capacity measurements on natural \chem{NaFeSi_{2}O_{6}} show that very little entropy is associated with the two previously identified magnetic transitions and instead short-ranged magnetic correlations build up in the quasi-one-dimensional chains over a temperature range extending well above $T_{\rm N}$. The $\mu$SR data appear similar to those for the synthetic sample, except that the oscillations have become incoherent. This suggests that the impurities change the magnetic ground state, either by breaking up the intra-chain ordering~\cite{imry75} or by inducing a significant staggered magnetization around the impurity sites~\cite{eggert95}. Because of the complexity of this natural system it is not possible to separate these possibilities, nor be certain which of the two phases observed in the natural sample is present in the synthetic sample. 
From the shapes of the hump due to short-ranged order in each sample we can estimate~\cite{dejongh74} intrachain exchange constants ($J^{\rm Li} \sim 7$~K, $J^{\rm Na} \sim 6.5$~K) roughly consistent with the calculations of \textcite{streltsov08}, though the comparison is complicated significantly by the interchain exchange.

We can also compare our results to those reported on the other quasi-one-dimensional multiferroics \chem{LiCu_{2}O_{2}}~\cite{rusydi08} and \chem{Ca_{3}(Co,Mn)_{2}O_6}~\cite{choi08}. The analogy with \chem{LiCu_{2}O_{2}} is somewhat closer, particularly for natural \chem{NaFeSi_{2}O_{6}}, since there are two closely spaced magnetic transitions bounding a magnetically but not ferroelectrically ordered intermediate phase. Our heat capacity measurements suggest that natural \chem{NaFeSi_{2}O_{6}} has considerably more one-dimensional magnetic interactions than \chem{LiCu_{2}O_{2}}, but the magnetic structure is likely to be similar. \chem{Ca_{3}(Co,Mn)_{2}O_{6}} has a similar hump in its heat capacity but below the onset of ferroelectricity and without pronounced features.~\cite{choi08} 
Both \chem{LiCu_{2}O_{2}} and \chem{Ca_{3}(Co,Mn)_{2}O_6} show considerable evidence for disorder influencing the multiferroic properties, due to \chem{Li} non-stoichiometry and on-chain site disorder respectively. Comparison with the $\mu$SR data for \chem{Ca_{3}(Co,Mn)_{2}O_6}~\cite{lancaster09} shows that natural \chem{NaFeSi_{2}O_{6}} may indeed show some influence from incommensurate magnetism, on-chain site disorder, and correlations along the chains, as the muon precession we expect for the ordered states is evidently incoherent.

\section{\label{sec:conc} Conclusions}
In conclusion, we have investigated the magnetic properties of \chem{LiFeSi_{2}O_{6}} and both natural and synthetic \chem{NaFeSi_{2}O_{6}} using muon spin relaxation and heat capacity measurements. \chem{LiFeSi_{2}O_{6}} enters a commensurate antiferromagnetic state below $T_{\rm N} = 18.5$~K whereas both \chem{NaFeSi_{2}O_{6}} samples appear to be incommensurate. In the natural sample, impurities within the \chem{Fe} chains disturb this state and no coherent muon oscillations are observed. An unusual hysteresis is apparent in applied field heat capacity measurements of \chem{LiFeSi_{2}O_{6}} suggesting that magnetic correlations can be locked in well above $T_{\rm N}$. Two magnetic transitions are apparent in the heat capacity of the natural \chem{NaFeSi_{2}O_{6}} but no related features occur in the $\mu$SR data, which show a gradual build-up of static magnetism with decreasing temperature, or in the measurements on the synthetic sample, where only one transition is evident. Investigations of impurity effects in model one-dimensional chain magnets could shed considerable light on this enigmatic behavior. Further work would be worthwhile to make synthetic single crystals of \chem{NaFeSi_{2}O_{6}} suitable for dielectric measurements to clarify whether the multiferroicity is indeed being triggered by disorder, and also to determine the magnetic structures of both synthetic and natural \chem{NaFeSi_{2}O_{6}} to gain a better understanding of the multiferroic mechanism in this compound.

\acknowledgments
Part of this work was performed at the Swiss Muon Source, Paul Scherrer Institute, Villigen, CH. We are grateful to Hubertus Luetkens for experimental assistance, Pierre Toledano for helpful discussions, and to the EPSRC and STFC (UK) for financial support. This research project has been supported by the European Commission under the 7th Framework Programme through the `Research Infrastructures' action of the `Capacities' Programme, Contract No: CP-CSA\_INFRA-2008-1.1.1 Number 226507-NMI3.

\bibliography{pjb-fso}

\end{document}